\documentclass[english, a4paper, 12pt]{article}
\usepackage{geometry}
\usepackage{graphicx}
\usepackage{amsmath}
\usepackage{amssymb}
\usepackage{multicol}
\usepackage{algorithmicx}
\usepackage{algpseudocode}
\usepackage{algorithm}
\usepackage{color}
\usepackage{natbib}
\usepackage{setspace}
\usepackage{url}
\usepackage{lineno}

\newcommand{\bY}{\mathbf{Y}}
\newcommand{\bX}{\mathbf{X}}
\newcommand{\bZ}{\mathbf{Z}}
\newcommand{\bD}{\mathbf{D}}
\newcommand{\bu}{\mathbf{u}}
\newcommand{\bLx}{\mathbf{L_X}}
\newcommand{\balpha}{\boldsymbol{\alpha}}
\newcommand{\btheta}{\boldsymbol{\theta}}
\newcommand{\bmu}{\boldsymbol{\mu}}
\newcommand{\btau}{\boldsymbol{\tau}}
\newcommand{\bgamma}{\boldsymbol{\gamma}}

\begin{document}

%%%%%%%%%%%%%%%%%%%%%%%%%%%%%%%%%%%%%%%%%%%%%%%%%%%%%%%%%%%%%%%%%%%%%%%%%%%%%%
%%%%%%%%%%%%%%%%%%%%%%%%%%%%%%%%%%%%%%%%%%%%%%%%%%%%%%%%%%%%%%%%%%%%%%%%%%%%%%
%%%%%%%%%%%%%%%%%%%%%%%%%%%%%%%%%%%%%%%%%%%%%%%%%%%%%%%%%%%%%%%%%%%%%%%%%%%%%%
%\thispagestyle{empty}
\begin{center}
{\bf \Large Spatially-constrained clustering of ecological networks}

\vspace{0.5cm}
Vincent Miele\footnote{Corresponding author: \texttt{vincent.miele@univ-lyon1.fr}}, Franck Picard and St\' ephane Dray

\vspace{0.5cm}
Laboratoire Biom\'etrie et Biologie Evolutive,
Universit\'e  de Lyon,\\Universit\'e Lyon 1, CNRS, UMR5558; Villeurbanne, France.\\

\end{center}
%\newpage
%\linenumbers

%%%%%%%%%%%%%%%%%%%%%%%%%%%%%%%%%%%%%%%%%%%%%%%%%%%%%%%%%%%%%%%%%%%%%%%%%%%%%%
%%%%%%%%%%%%%%%%%%%%%%%%%%%%%%%%%%%%%%%%%%%%%%%%%%%%%%%%%%%%%%%%%%%%%%%%%%%%%%
%%%%%%%%%%%%%%%%%%%%%%%%%%%%%%%%%%%%%%%%%%%%%%%%%%%%%%%%%%%%%%%%%%%%%%%%%%%%%%
{\parindent0pt
{\bf Abstract}\\
Spatial ecological networks are widely used to model interactions between georeferenced biological entities (e.g., populations or communities). The analysis of such data often leads to a two-step approach where groups containing similar biological entities are firstly identified and the spatial information is used afterwards to improve the ecological interpretation.
We develop an integrative approach to retrieve groups of nodes that are geographically close and ecologically similar. Our model-based spatially-constrained method embeds the geographical information within a regularization framework by adding some constraints to the maximum likelihood estimation of parameters.
A simulation study and the analysis of real data demonstrate that our approach is able to detect complex spatial patterns that are ecologically meaningful.
The model-based framework allows us to consider external information (e.g., geographic proximities, covariates) in the analysis of ecological networks and appears to be an appealing alternative to consider such data.\\

\vspace{0.5cm}
}

%\begin{multicols}{2}
%%%%%%%%%%%%%%%%%%%%%%%%%%%%%%%%%%%%%%%%%%%%%%%%%%%%%%%%%%%%%%%%%%%%%%%%%%%%%%
%%%%%%%%%%%%%%%%%%%%%%%%%%%%%%%%%%%%%%%%%%%%%%%%%%%%%%%%%%%%%%%%%%%%%%%%%%%%%%
%%%%%%%%%%%%%%%%%%%%%%%%%%%%%%%%%%%%%%%%%%%%%%%%%%%%%%%%%%%%%%%%%%%%%%%%%%%%%%
\section*{Introduction}

In many ecological studies, researchers must analyze data describing the interactions between biological entities (e.g., individuals, populations, species or communities). These interactions can be directly observed \citep[e.g., trophic relationships in a food-web,][]{Krause2003} or they can be inferred from computed distance/similarity measures. Several genetic distances have been developed to summarize allele frequency differences between populations for instance \citep{Kalinowski2002}. In community ecology, species abundance (or presence/absence) data are routinely used to assess turnover (i.e., beta-diversity) between sites based on differences in species composition \citep[Chapter 7 in][]{Legendre2012a}. Describing and summarizing these sets of pairwise interactions is an important step to better understand the functioning of ecological systems. In a theoretical viewpoint, networks (or graphs) offer a natural and efficient framework to store and analyze interaction data. Whereas food-web analysis was a traditional and historical field \citep{Ings2009}, network analysis techniques have recently gained popularity in genetics \citep{kel13,alb13}, movement ecology  \citep{jac12}, landscape ecology \citep{Bodin2006, per11}, biogeography \citep{Thebault2013, moa12} or species distribution modeling \citep{Foltete2012, ara11}. 

From a statistical perspective, interaction data can be handled using \textit{ecological networks} where biological entities correspond to the nodes while the intensities of interactions are represented by weighted edges \citep{Proulx2005}. Traditional approaches consist in summarizing the structure of a network using easy-to-compute statistics that relate to ecological properties (connectivity, degree distribution, average path length, \cite{ray11}). Unsupervised clustering is another common practice to detect modules {\it i.e.} groups of biological entities more densely connected to each other than to other external entities. In ecology, this strategy has been widely applied: several works have linked the modularity to the stability of the ecological network \citep{Krause2003}; finding modules has also helped to identify co-evolution patterns \citep{Dupont2009}, to delineate conservation units \citep{Fortuna2009}, habitat patches \citep{per11} or biogeographic entities \citep{moa12}.

The identification of groups of nodes (such as modules for instance) has received considerable interest in physical sciences (see \citet{Newman2006} or \citet{fort10} for an extensive review) but also in statistics: in particular this work relies on well-established model-based clustering procedures \citep{dau08,pic09}. These methods are fundamentally different from classic approaches by assuming a statistical distribution for interaction data. They are thus very flexible for handling various types of interaction data by specifying adequate statistical distribution to model presence/absence (Bernoulli), abundances (Poisson) or fluxes (Gaussian) or to integrate external covariates \citep{mar10}. This general statistical framework is adequate for using tools from  model selection theory (such as Bayesian Information Criterion (BIC)) in order to determine the number of groups that structure the data. Lastly, these methods can detect hidden connectivity patterns that are not limited to modularity, 
such as centrality (hubs) or hierarchy (see \cite{pic09} on a food web of wasps).

In many situations, the nodes correspond to entities that have explicit geographic locations transforming ecological networks into \textit{spatial networks} \citep{dal10}. When available, this spatial information is often used \textit{posterior} to the identification of modules to improve their ecological interpretation \citep{moa12,ara11, Dattilo2013}. However, if the aim of a study is to identify spatially-coherent modules (e.g., habitat patches), this indirect approach may not be optimal as it considers the spatial aspect only after summarizing the network structure. An appropriate technique would integrate spatial information explicitly in the detection of modules. This process is the core of spatially-constrained clustering techniques \citep{amb98,Gordon1996, Duque2007} when ecological observations are stored as raw data tables. When data consists in pairwise interactions stored as a network, no direct technique yet exists to delimit spatial modules except related strategies based on boundary detection \citep{Monmonier1973}.

In this work, we propose an original method to integrate the spatial information in the clustering of ecological interactions. Whereas spatial information is traditionally treated as geographic coordinates, trend surfaces or distance matrix, we used a more efficient strategy that models spatial proximity as a {\it structural network} where two nodes are connected if they are considered as neighbors \citep{Dray2006, Dray2012}. This approach offers a great flexibility allowing to include the effect of structural constraints (landscape fragmentation) or physical barriers (rivers, mountains). In our framework, what is usually refered to a spatial network is thus decomposed into a pair of networks:  nodes are identical but edges reflect either biological interactions (ecological network) or spatial connectivity (structural network). Our new model-based approach considers both networks in a single step allowing to identify spatially-coherent modules. The originality of our method is to embed the geographical information within a regularization framework \citep{amb98,he11} that has been popularized by the analysis of high dimensional data sets \citep{has2001,buhl2011}. We evaluate our new approach by a simulation study and applied it on a real ecological data set.
An R package including methods and data to perform the analysis is also provided at \url{http://lbbe.univ-lyon1.fr/geoclust}.

%%%%%%%%%%%%%%%%%%%%%%%%%%%%%%%%%%%%%%%%%%%%%%%%%%%%%%%%%%%%%%%%%%%%%%%%%%%%%%
%%%%%%%%%%%%%%%%%%%%%%%%%%%%%%%%%%%%%%%%%%%%%%%%%%%%%%%%%%%%%%%%%%%%%%%%%%%%%%
%%%%%%%%%%%%%%%%%%%%%%%%%%%%%%%%%%%%%%%%%%%%%%%%%%%%%%%%%%%%%%%%%%%%%%%%%%%%%%
\section*{Materials and Methods}

\subsection*{Model-based Clustering of Interaction Data}
Our method belongs to the general framework of model-based clustering of network data. This family of models includes the stochastic block model \citep{air08} and the MixNet approach \citep{dau08,mar10} among others. We consider data for $n$ interacting entities with $Y_{ij}$ standing for the observed measure of these interactions and $\bY=(Y_{ij})$. Our model is based on group-membership of entities and $\bZ=(\bZ_1,...,\bZ_n)$ denotes the matrix of labels of entities $1, .., n$, {\it i.e.} $Z_{iq}=1$ if $i$ belongs to group $q$ and $0$ otherwise. In the context of unsupervised clustering, this matrix is unknown and our method aims to recover these labels using the observed information contained in $\bY$. Model-based clustering hypotheses that if labels were known, the distribution of the interaction data would be completely determined. Hence, we start by assuming that there are $Q$ groups with proportions $\balpha = (\alpha_1, ..., \alpha_Q)$ such that the distribution of labels $\bZ_i=(Z_{i1},...Z_{iQ})$ is Multinomial with parameter $\balpha$. The number of groups $Q$ is unknown and will be estimated afterwards. The distribution of the interaction data is specified conditionally to the labels:
\begin{equation}
\bZ_i \sim \mathcal{M}(1,\balpha), \;\; \bZ_j\sim \mathcal{M}(1,\balpha), \;\; Y_{ij}| \{Z_{iq}Z_{i\ell} = 1 \} \sim f(\cdot,\theta_{ql}),
\end{equation}
where distribution $f(\cdot,\theta_{ql})$ can be Bernoulli to model presence-absence data \citep{air08,dau08}, Gaussian or Poisson to model fluxes or abundance data \citep{mar10}. The parameters of this model are the proportions for the groups ($\balpha$) and the parameters governing the conditional distribution of the observations ($\btheta=(\theta_{q \ell})$). In the following, we note  $\bgamma=(\balpha,\btheta)$. \\

The objective is to estimate $\bgamma$ and to recover the unobserved labels of the data using the posterior expectation of membership $\mathbb{E}(\bZ_i|\bY)$. This is achieved using the EM-algorithm to maximize the observed-data likelihood denoted by $\log \mathcal{L}(\bY;\bgamma)$. Unfortunately, the direct maximization of this likelihood is untractable due to the total number of possible partitions $(\mathcal{L}(\bY;\bgamma)=\prod_{\bZ} \mathcal{L}(\bY,\bZ;\bgamma))$. Hence, we use an iterative algorithm that maximizes the complete-data likelihood: 
\begin{eqnarray}
\log \mathcal{L}(\bY,\bZ;\bgamma) & = & \log \mathcal{L}(\bZ;\balpha)+\log \mathcal{L}(\bY|\bZ;\btheta) \nonumber \\
&=& \sum_{iq} Z_{iq} \log(\alpha_q) + \sum_{ij,q\ell}Z_{iq}Z_{j\ell} \log f(Y_{ij}; \theta_{q \ell})
\end{eqnarray}
The labels being unknown, the algorithm proceeds as follows: the E-Step  
computes the conditional expectation of the complete-data log-likelihood defined as:
\begin{eqnarray}
\mathcal{Q}(\bgamma, \bgamma^{[h]}) &=& \mathbb{E}_{\bgamma^{[h]}} \left\{\log \mathcal{L}(\bY,\bZ;\bgamma)|\bY\right\} \nonumber  \\
&=& \sum_{iq} \mathbb{E}_{\bgamma^{[h]}}(Z_{iq}|\bY) \log(\alpha_q) + \sum_{ij,q\ell} \mathbb{E}_{\bgamma^{[h]}}(Z_{iq}Z_{j\ell}|\bY) \log f(Y_{ij}; \theta_{q \ell}),
\end{eqnarray}
for a current value of the parameters $(\bgamma^{[h]})$. Then the M-step maximizes $\mathcal{Q}$ with respect to $\balpha$ and $\btheta$. Computational difficulties often arise at the E-step, mainly due to complex dependency structures that can govern the $posterior$ distribution of labels given the data. This issue has motivated many methodological developments, in particular in the context of network data, with the use of variational methods \citep{jor99} to compute this $posterior$ distribution \citep{dau08,mar10}. 

%%%%%%%%%%%%%%%%%%%%%%%%%%%%%%%%%%%%%%%%%%%%%%%%%%%%%%%%%%%%%%%%%%%%%%%%%%%%%%
%%%%%%%%%%%%%%%%%%%%%%%%%%%%%%%%%%%%%%%%%%%%%%%%%%%%%%%%%%%%%%%%%%%%%%%%%%%%%%
\subsection*{Accounting for Spatial Constraints in the Clustering Model} 

\subsubsection*{Labels regularization using a spatial network}

\begin{figure}[h!]
  \begin{center} 
    \includegraphics[height=19cm]{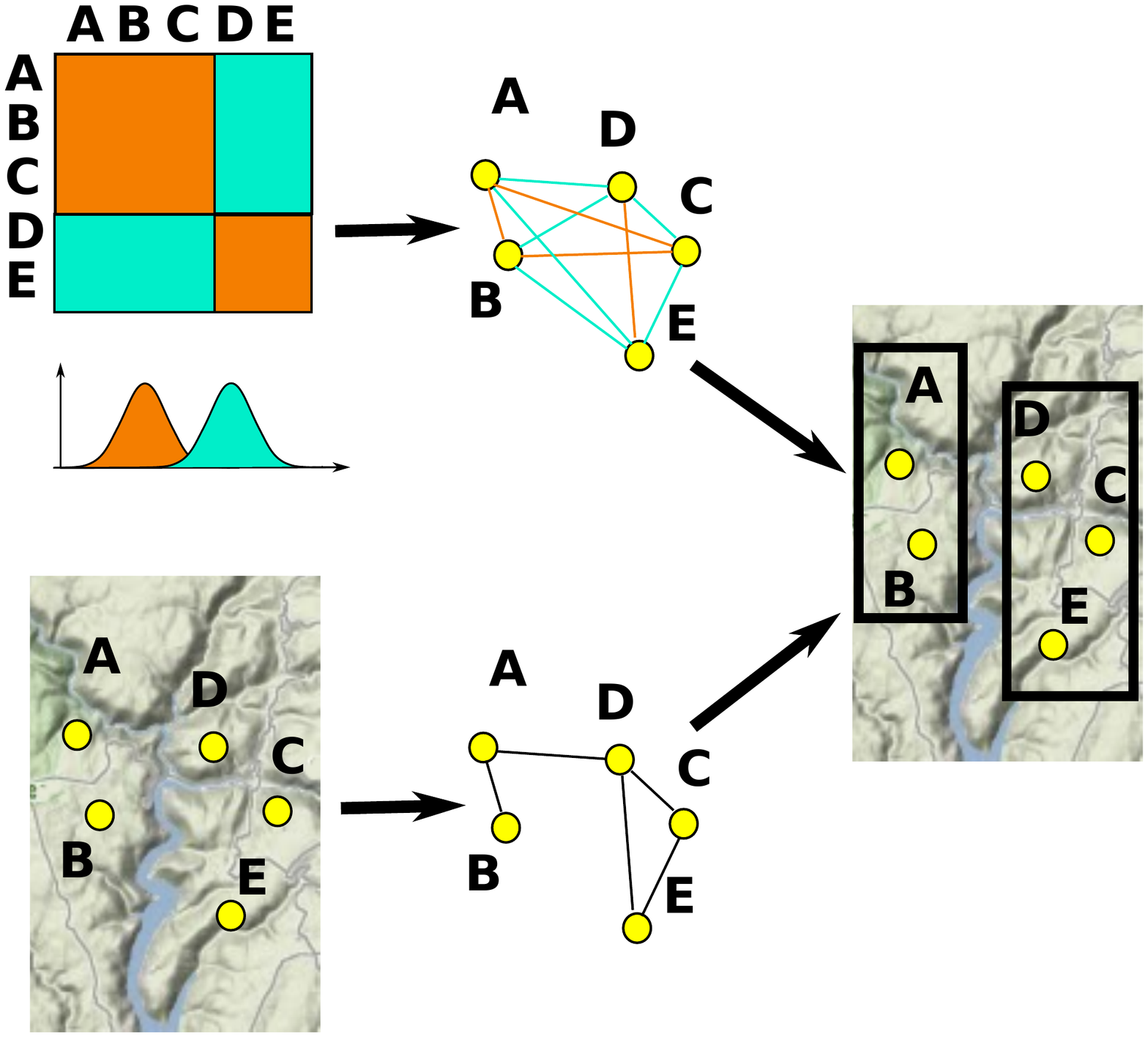}
    \caption{Framework of our algorithm. The ecological network (top) records the ecological distance between entities.
The structural network (bottom) summarizes the proximity between geographical locations. 
Our method deciphers groups of entities (black squares on the right) using both networks into a model-based strategy associated to a regularization framework.
While entities A,B,C and D,E forms two groups in the ecological data, the geographical constraints leads to lastly grouping C with D and E.}    	
  \label{figframework}
  \end{center}
\end{figure}

We choose to use a structural network that records the spatial proximity between ecological entities of the ecological network $\bY$ (see Figure \ref{figframework}). Structural networks are sometimes directly available such as road networks \citep{sma12}, but they are usually constructed using geographical data with {\it ad-hoc} techniques such as  maximum spanning trees \citep{ass06}, k-nearest neighbors \citep{guo08}, distance threshodling \citep{per11} or edge-thinning \citep{urb01, kel13}. In the following we suppose that the structural network is given and fixed. It is denoted by $\bX=(X_{ij})$ such that $(X_{ij}>0)$ is the geographical proximity between entities $i$ and $j$  and $X_{ij}=0$ if they are not connected. The entities are the same as in the ecological network $\bY$.

We propose to embed the geographical information within a regularization framework by adding some constraints in the maximum likelihood estimation of parameters. In regularization techniques, a constraint defined by a network can be introduced using the graph Laplacian \citep{jacob12}. For a network with connection matrix $\bX=(X_{ij})$, the Laplacian is defined by $\bLx=\bD-\bX$ where $\bD$ is the diagonal matrix of degrees with diagonal terms $d_{i} = \sum_{j}{X_{ij}}$. The Laplacian $\bLx$ can then be used as a metric to measure the spatial variability. Indeed, for a given vector $\bu=(u_1,...u_n)$, we have:
$$
\| \bu\|^2_{\bLx}= \bu^T \bLx \bu = \sum_{i j} X_{ij}(u_i-u_j)^2,
$$
which is the squared distance between values of $\bu$ weighted by their spatial proximities contained in $\bX$. This quantity has been defined as the local variance by \citet{Lebart1969} and is equal to the numerator of the spatial autocorrelation index proposed by \citet{Geary1954}.

We develop an original regularization procedure aiming to reduce the variation of labels along the spatial network. Whereas the vector of parameters is traditionally regularized, our approach considers that the vector of labels can be regularized using the spatial network $\bX$. Denoting by $\bZ^q=(Z_{1q},\hdots,Z_{nq})$ the vector of individuals for label $q$, we propose the following penalty:
$$
\text{pen}(\bZ;\bLx) = \sum_{q=1}^Q\| \bZ^q\|^2_{\bLx} =  \sum_{q=1}^Q \sum_{i,j} X_{ij}(Z_{iq}-Z_{jq})^2, 
$$
and the likelihood to maximize by the EM-algorithm becomes:
$$
\log \mathcal{L}(\bY,\bZ;\bgamma) - \lambda \times \text{pen}(\bZ;\bLx),
$$
with $\lambda$ a constant controlling the amount of penalization that can be estimated adaptively to the data. Let us consider the case where $X_{ij}\in \{0,1\}$ to 
interpret the penalty. In this case, 
$$
\text{pen}(\bZ;\bLx) =  \sum_{q=1}^Q \sum_{i \sim j}(Z_{iq}-Z_{jq})^2 = \sum_{q=1}^Q \sum_{i \sim j} 1_{\{Z_{iq} \neq Z_{jq}\}}
$$
with $i \sim j$ standing for entities $i$ and $j$ connected in the spatial network, so that the penalty accounts for the number of edges in the spatial network that have discordant labels. 

\subsubsection*{Regularized EM-algorithm based on spatial network}

Considering the new penalized likelihood, the regularized EM algorithm is based on the conditional expectation of the penalized complete-data likelihood
$$
\mathcal{Q}(\bgamma, \bgamma^{[h]}) - \lambda \times \mathbb{E}_{\bgamma^{[h]}} \left\{\text{pen}(\bZ;\bLx)|\bY\right\}.
$$
As the penalty term does not involve the parameters but only the labels, the maximization step is unchanged \citep{mar10}. For instance, if the interaction data are supposed to be Gaussian such that
\begin{equation}
\label{model-simu}
Y_{ij}| \{Z_{iq}Z_{i\ell} = 1 \} \sim \mathcal{N}(\mu_{q \ell}, \sigma^2),
\end{equation}
we get the following updated estimates (at iteration $[h+1]$):
\begin{equation*}
\widehat{\mu}_{ql}^{[h+1]}  =  \frac{\sum_{ij,q\ell}\widehat{Z}_{iq}^{[h]} \widehat{Z}_{j\ell}^{[h]} Y_{ij}}{\sum_{ij,q\ell}\widehat{Z}_{iq}^{[h]} \widehat{Z}_{j\ell}^{[h]}}, \; \;
\widehat{\sigma}^{2[h+1]}  =  \frac{\sum_{ij,q\ell} \widehat{Z}_{iq}^{[h]} \widehat{Z}_{jl}^{[h]} (Y_{ij} - \widehat{\mu}_{ql}^{[h+1]})^2}{\sum_{ij,q\ell}\widehat{Z}_{iq}^{[h]} \widehat{Z}_{j\ell}^{[h]}},
\end{equation*}
with $(\widehat{Z}_{iq}^{[h]})$ the predicted labels provided by the E-step (below).

\paragraph{Network-Based Regularized E-Step.} The traditional EM algorithm is not tractable in the case of mixtures for interaction data \citep{dau08}. Thus, we use a variational approach which consists in choosing a surrogate $posterior$ distribution for the labels so that we can suppose that they are conditionally independent \citep{jor99}. Denoting by $\tau_{iq} \simeq \mathbb{E}_{\bgamma^{[h]}} \left\{Z_{iq}|\bY\right\}$ the approximate posterior expectation of labels and by $\btau=(\btau_i)=(\tau_{iq})$, the penalized complete-data likelihood becomes:
\begin{eqnarray*}
\mathcal{Q}(\bgamma, \bgamma^{[h]}) - \lambda \times  \mathbb{E}_{\bgamma^{[h]}} \left\{\text{pen}(\bZ;\bLx)|\bY\right\} & \simeq & \sum_{iq} \tau_{iq} \log(\alpha_q) + \sum_{ij,q\ell} \tau_{iq} \tau_{j \ell} \log f(Y_{ij}; \theta_{q \ell}) \\
&-& \lambda \times \sum_{i,j} X_{ij}(\btau_i-\btau_j)^2.
\end{eqnarray*}
The computation of these approximate posterior probabilities is achieved by solving a fixed-point algorithm \citep{amb98,mar10} and we account for spatial constraints through the penalty term. The derivation of this algorithm is provided in the Appendix. Lastly, to retrieve clear and separable groups, we add a classification step corresponding to the Classification-EM algorithm \citep{cel92}. We use the Maximum \textit{a  posteriori} rule such that $\widehat{Z}_{iq} = 1$ if $q =  \operatorname*{arg\,max}_{\ell}  \widehat{\tau}_{i\ell}$ and $0$ otherwise.

\paragraph{Initialization, $\lambda$  values and number of groups}
It is well know that EM-algorithms are very sensitive to the quality of the initialization point. We propose to set-up $\btau^{(0)}$ using the clustering partition $\widehat{\bZ}^0$ obtained with a traditional k-means algorithm as suggested by \citet{dau08}. Parameter $\lambda$ increased until $\widehat{\lambda}_{\max}$ that corresponds to the maximal spatial homogeneity (i.e. until no further change of labels). Lastly, the number of groups $\widehat{Q}(\widehat{\lambda}_{\max})$ is chosen using model selection strategy based on the Integrated Classification Likelihood (ICL, \cite{dau08}).

%%%%%%%%%%%%%%%%%%%%%%%%%%%%%%%%%%%%%%%%%%%%%%%%%%%%%%%%%%%%%%%%%%%%%%%%%%%%%%
%%%%%%%%%%%%%%%%%%%%%%%%%%%%%%%%%%%%%%%%%%%%%%%%%%%%%%%%%%%%%%%%%%%%%%%%%%%%%%
%%%%%%%%%%%%%%%%%%%%%%%%%%%%%%%%%%%%%%%%%%%%%%%%%%%%%%%%%%%%%%%%%%%%%%%%%%%%%%
\section*{Results}
\subsection*{Simulation study}

In this section we evaluate the relative merits of our method on simulated data that explore different degrees of coherence 
between the interaction data and the spatial information.
For standard clustering methods, the performance only depends on the {\it group separability}, 
{\it i.e.} the ability to delineate groups within the ecological network.
However, in the case of spatially constrained clustering, performance is more difficult to explore 
 since it depends on a trade-off between group separability and spatial homogeneity/coherence of the groups. 
The easiest configuration is when the groups and the spatial information overlap (low {\it spatial discordance}), 
the most difficult situation is when groups and spatial information are  contradictory (high {\it spatial discordance}). 

We start by simulating a structural network $\bX$ that consists of two spatial components of 50 nodes each, related by one single edge only (see Supplementary Figure 1). 
The proximity values stored in the  edges in $\bX$ are determined using a Gabriel graph \citep{dal10}.
Then we consider $Q=2$ groups with equal proportions $\balpha^*=(0.5,0.5)$, 
and we sample the true labels $\bZ^*$ such that $\bZ_{i} \sim \mathcal{M}(1, \balpha^*)$. 
In order to explore different degrees of spatial discordance for the groups, we start by matching labels with the connected components of $\bX$ (no spatial discordance), 
and we randomly sample an increasing number of pairs of nodes from the two spatial components and swap their labels  (without modifying $\bX$). 
Then we use the proportion of edges in $\bX$ with discordant labels in the spatial network as a measure of spatial discordance. 
In the most favorable situation, there is a perfect overlap between labels and the spatial components in $\bX$. 
In the worst situation, labels are spread independently of $\bX$. 
Once the spatial discordance of labels is determined, we sample a Gaussian interaction network $\bY$ 
(Eq. \ref{model-simu}) with parameters $\sigma=0.2$ and mean parameters:
$$\bmu=  \left[\begin{array}{cc}\mu & \nu \\ \nu & \mu\end{array}\right].$$
The group separability  is governed by $\delta = (\mu-\nu)/\sigma$ which varied from no signal ($\delta=0$) to high signal levels ($\delta=1$).
Each configuration was repeated 50 times and we applied our method with spatial penalty or without ({\it i.e.} $\lambda=0$). 
The performance of the method is then assessed by the averaged adjusted Rand Index (aRI, \cite{hub85}) that measures the agreement between two partitions. 
It lies between 0 and 1, 1 being for a perfect match between the simulated and the recovered partition. 
This index is first used to assess the classification of interactions (between estimated and simulated labels $\bZ^*$, referred to as {\it aRI on $\bY$}, with the true classification given by the two spatial components), 
and secondly to assess the spatial homogeneity of groups (between estimated labels and the two spatial components of $\bX$, referred to as {\it aRI on $\bX$}).

\begin{figure}[h!]
  \begin{center}
    \includegraphics[width=14.5cm]{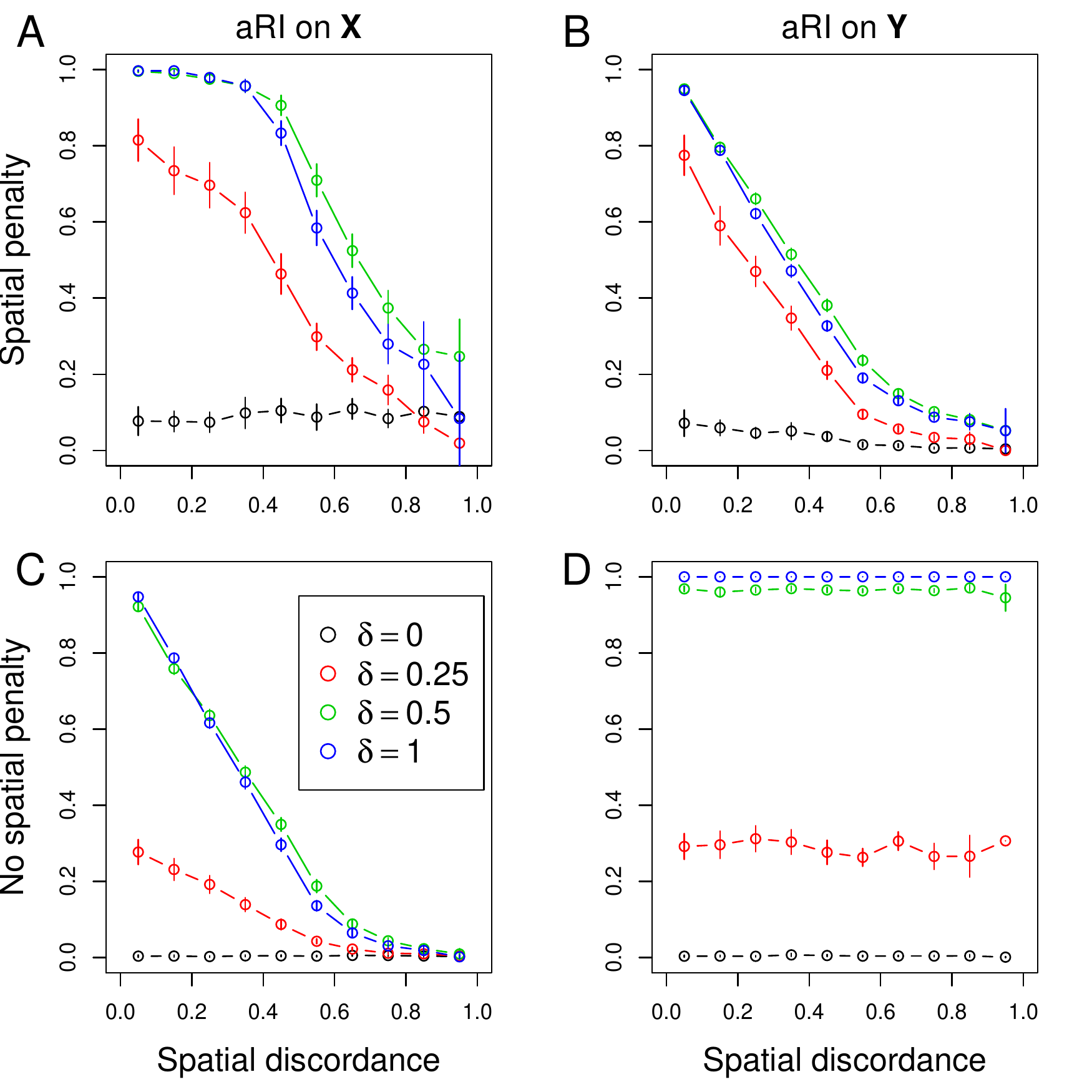}
    \caption{Performance of our method on simulated data, assessed by the averaged Adjusted Rand Index (aRI), 
      for varying spatial discordance and group separability ($\delta$) values. 
      (A) aRI on $\bX$ with our regularized EM algorithm with spatial penalty. (B) same for aRI on $\bY$.
      (C) aRI on $\bX$ without spatial penalty, {\it i.e} $\lambda=0$. (D) same for aRI on $\bY$.}
    \label{figsimuplot}
  \end{center}
\end{figure}

When groups are highly separable ($\delta \geq 0.5$) and without any spatial discordance, clustering performs perfectly with or without the spatial penalty (aRI $\simeq$ 1 on both $\bX$ and $\bY$, Figure \ref{figsimuplot}A-D). The interest in using the spatial penalty is illustrated when the spatial discordance increases until the proportion of spatially discordant labels reaches 50\%. In this case, the spatial penalty induces a lower classification performance on the interaction data (decreasing aRI on $\bY$, Figure \ref{figsimuplot}B), 
which is compensated by an excellent clustering performance on spatial information (aRI$>$0.9 on $\bX$, Figure \ref{figsimuplot}A).
Consequently, the effect of the spatial penalty is to maintain the spatial homogeneity of the groups at the price of classification errors on $\bY$. 
When the separability of the group decreased ($\delta=0.25$), the performance of clustering also decreased, with a aRI of 0.3 on $\bY$. 
In this case, the spatial penalty clearly helps to recover the groups when they match the spatial structure 
(aRI=0.8 on $\bX$ without spatial discordance, $\delta=0.25$, Figure \ref{figsimuplot}A,C). 
In conclusion, when the spatial discordance is low to moderate, the spatial penalty helps in finding the groups, even if the signal in the interaction network is moderate.

\begin{figure}[h!]
  \begin{center}
    \includegraphics[width=7.5cm]{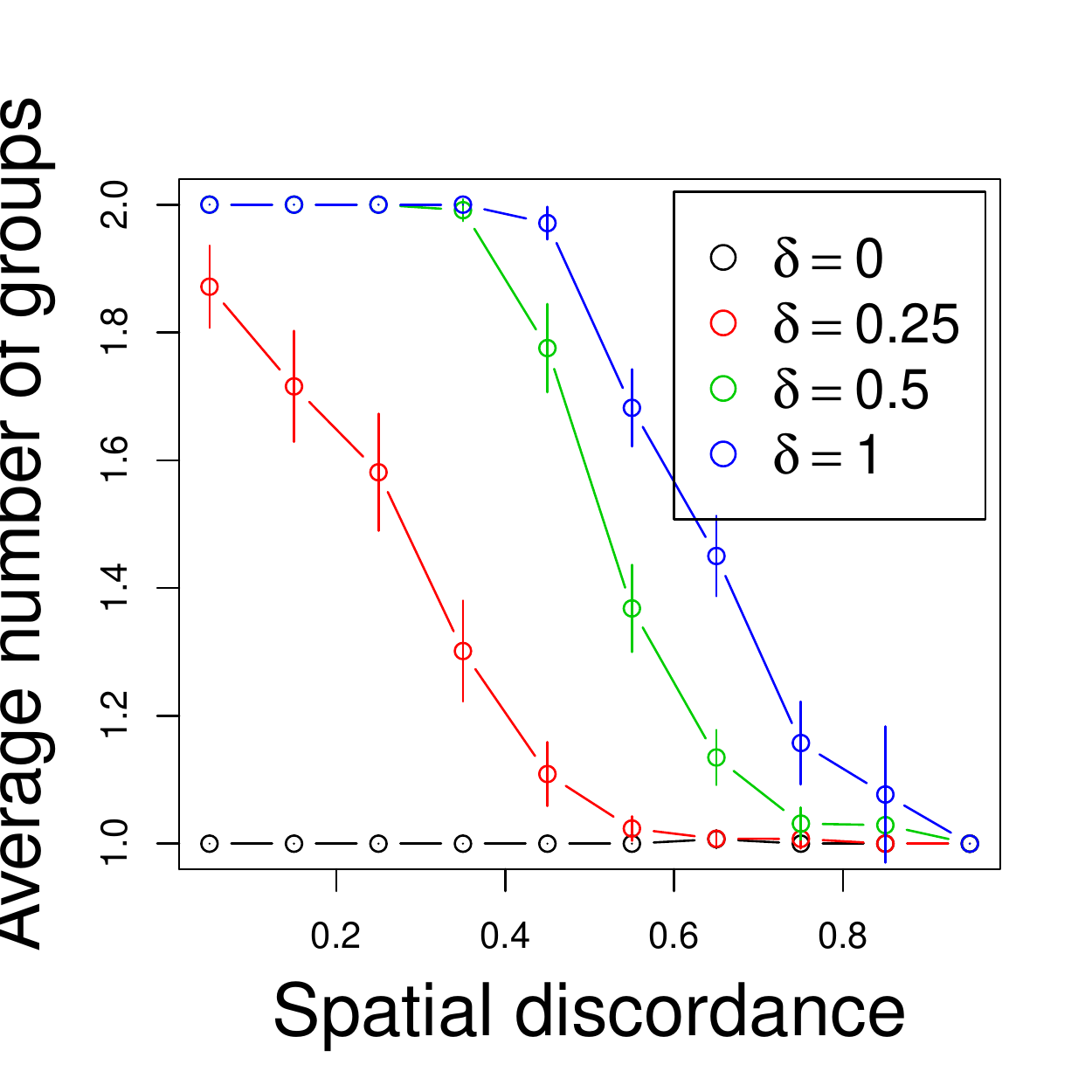}
    \caption{Average number of groups selected by our model selection approach, on simulated data, 
      for varying spatial discordance and group separability ($\delta$) values.
    }
    \label{figsimuplot2}
  \end{center}
\end{figure}

Lastly, we evaluate the ability of our method to identify groups when the spatial discordance is high. When looking at the number of groups chosen by the model selection procedure (Figure \ref{figsimuplot2}), it is clear that all nodes are gathered in a single group when the spatial discordance increases. This proves that our method is adaptive to the spatial discordance, and searches only for \textit{spatially-coherent modules}: when there is no spatial signal, our method returns no clustering instead to identify groups that are not spatially organized. 

\subsection*{Hydrothermal vents}

We re-examine data concerning the presence/absence of 332 genera in 63 hydrothermal fields. This data set was used by \citet{bac09} and \citet{moa12} to define biogeographic provinces based on the faunal distribution in oceanic hydrothermal vents. We built an ecological network by computing distances in faunal composition using Jaccard coefficient \citep[see Chapter 7 in][]{Legendre2012a}.  This fully connected network contains 63 nodes (the fields) and edges are weighted by the ecological distances. 

The distribution of Jaccard coefficients was bimodal due to an excess of distance values equal to 1 corresponding to pairs of fields sharing no species; 
the remaining values, after a logistic transform, follow a Gaussian distribution. Consequently, we modelled data using an inflated Gaussian distribution. 
We computed spatial proximities among the fields by computing great circle distances. The structural network was then built by removing all the edges corresponding to distances higher than 3600 km (see Figure \ref{figappli2}B). This choice ensures that each node has at least one edge and that only local spatial proximities are considered. Edges were then weighted by spatial distances ($max(d_{ij}) - d_{ij})$.

\begin{figure}[h!]
  \begin{center}
    \includegraphics[width=\textwidth]{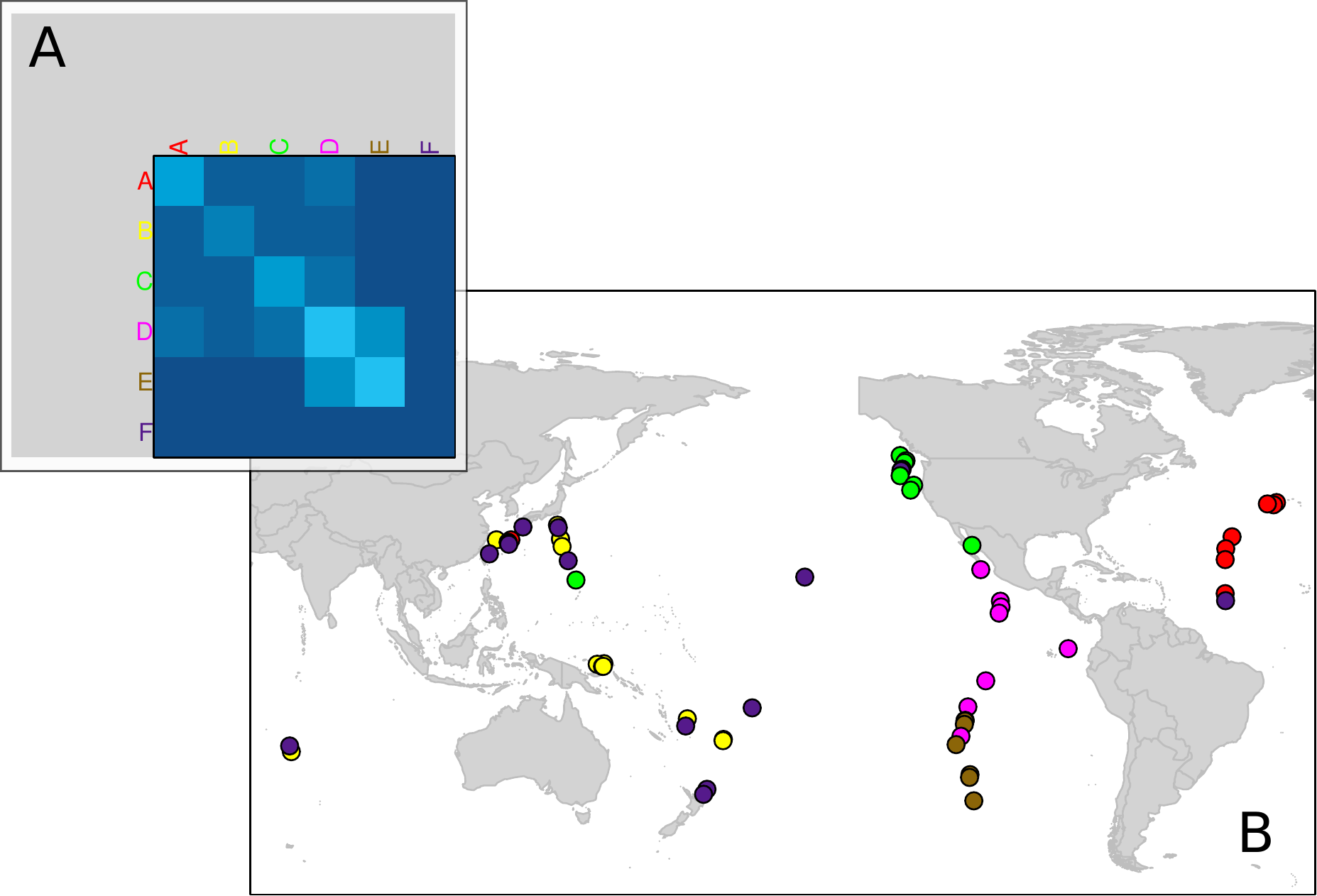}
    \caption{Clustering of 63 hydrothermal fields in 6 groups based on similarities in species composition (no spatial penalty). (A) Heatmap showing average ecological distances within and between groups. (B) Map showing the 6 groups}
    \label{figappli}
  \end{center}
\end{figure}

\begin{figure}[h!]
  \begin{center}
    \includegraphics[width=\textwidth]{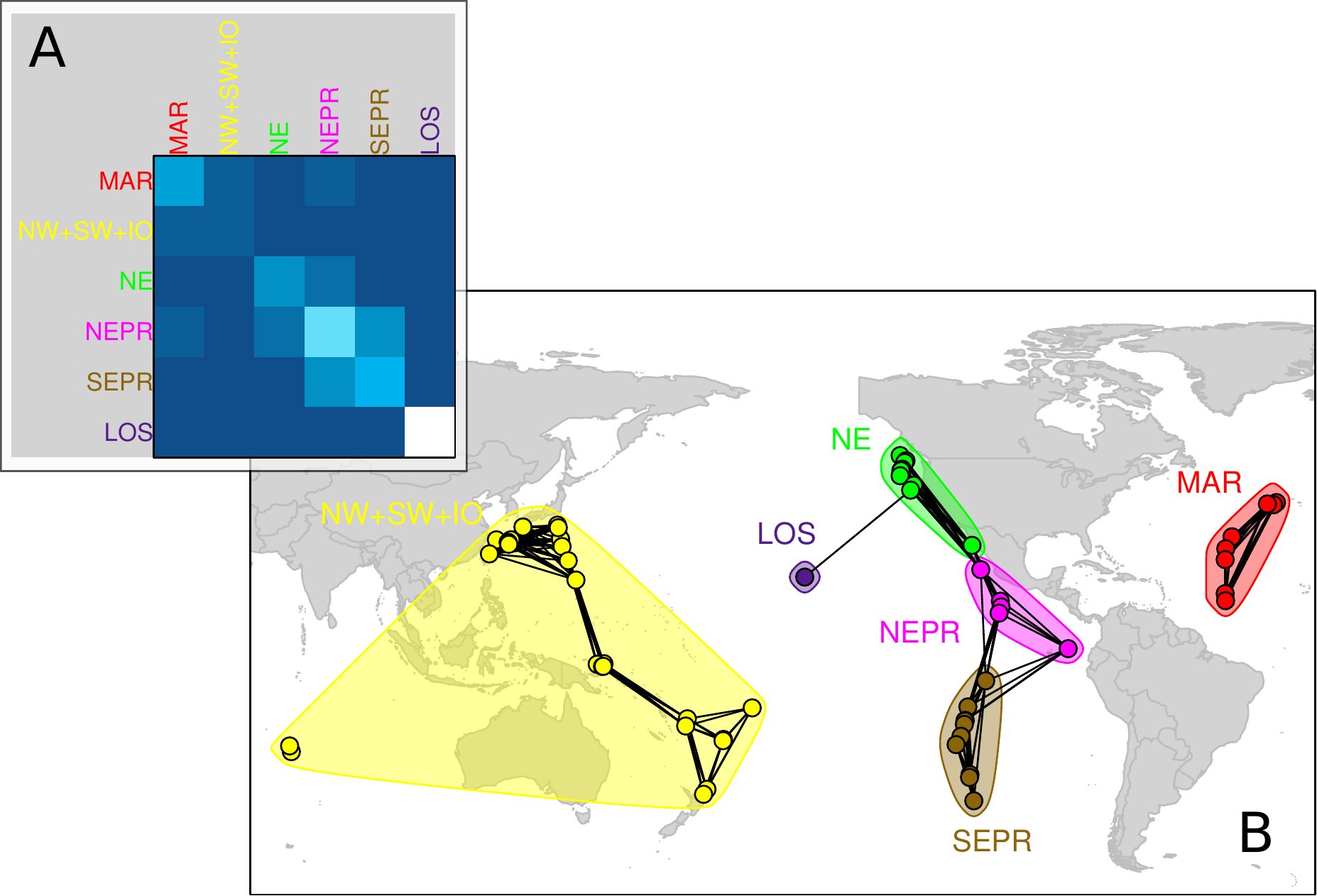}
    \caption{Spatial clustering of 63 hydrothermal fields in 6 groups based on similarities in species composition. (A) Heatmap showing average ecological distances within and between groups. (B) Map showing the structural network and the 6 groups: Garbage (GARB), NorthEast Pacific (NE), NorthEast Pacific Rise (NEPR), NorthEast Pacific Rise (SEPR), Mid-Atlantic Ridge (MAR) and NorthWest Pacific/Southwest Pacific/Indian Ocean (NW+SW+IO).}
    \label{figappli2}
  \end{center}
\end{figure}

We first applied our algorithm without spatial penalty (i.e., $\lambda = 0$). Fields were partitioned in 6 groups based on their species composition (Figure \ref{figappli}). It is clear that fields share more ecological similarities within groups that between groups (comparison of diagonal and off-diagonal terms respectively in Figure \ref{figappli}A). This partitioning highlights some geographical patterns: some groups are roughly spatially coherent (groups A, C, D, E) whereas other groups did not show any spatial coherence (Figure \ref{figappli}B). 

If the spatial penalty is applied, the model selection procedure selected also a clustering in 6 groups (Figure \ref{figappli2}B). In this case, within-group similarities are also higher than the between-group (Figure \ref{figappli2}A). Average within-group Jaccard index (0.944) is slightly higher than for the unconstrained approach (0.928). This loss of ecological homogeneity is counterbalanced by a gain in the spatial aspect (Figure \ref{figappli2}B) as all ecological groups are now spatially coherent. Our results are very similar to those obtained by \citet{bac09} and \citet{moa12} who used different methods and identified 6 and 5 provinces, respectively. Their partitioning disagreed for the provinces in West Pacific (WP), Indian Ocean (IO) and East Pacific Rise (EPR). We identified a group with only one field (Loihi Seamount, LOS) that share no species with all others and is poorly spatially connected (only one edge) in the structural network. As in the mentioned studies, we found the Mid-Atlantic Ridge (MAR) and the NorthEast Pacific (NE) provinces, the latter including the Guaymas Basin in our results. We also identified the partitioning between the Northern and Southern-Pacific Rise provinces (NEPR and SEPR) that was proposed in \citet{bac09} as centers of dispersal for hydrothermal fauna. This shows that our algorithm can separate groups of fields that are spatially connected but very dissimilar according to the ecological network. Our algorithm was not able to partition the global province (NW+SW+IO on Figure \ref{figappli2}B) composed by the Indian Ocean (IO) and Western Pacific zones. Interestingly, both previous studies were discordant for this region: \citet{bac09} merged Indian Ocean and Southwest Pacific (SW) and introduced an additional Northwest Pacific province (NW) whereas  \citet{moa12} proposed a global WP province but kept aside the IO province. \citet{moa12} focused on the central role of the WP province in the biogeography of hydrothermal vents and also mentioned that the existence of the IO province remains unclear due its poor representation in the data. Consequently, the group NW+SW+IO is the most ecologically heterogeneous in our results (Figure \ref{figappli2}A) because our method is more conservative and suggests that any partitioning would be artificial in this region. This merging of IO and Western Pacific zones (SW, NW) demonstrated that our algorithm is able to merge two groups that are ecologically similar but not spatially connected.

%%%%%%%%%%%%%%%%%%%%%%%%%%%%%%%%%%%%%%%%%%%%%%%%%%%%%%%%%%%%%%%%%%%%%%%%%%%%%%
%%%%%%%%%%%%%%%%%%%%%%%%%%%%%%%%%%%%%%%%%%%%%%%%%%%%%%%%%%%%%%%%%%%%%%%%%%%%%%
%%%%%%%%%%%%%%%%%%%%%%%%%%%%%%%%%%%%%%%%%%%%%%%%%%%%%%%%%%%%%%%%%%%%%%%%%%%%%%
\section*{Conclusions}
The analysis of ecological networks has gained a lot of attention in the past years, 
and the development of dedicated statistical methods has been challenging. 
Here we propose a statistical framework to identify spatial groups of ecological entities based on interaction data.
We developed an efficient way to retrieve spatially-coherent groups of ecological entities using a regularization framework.
The analysis of simulated data and of a real dataset demonstrated the ability of our method to identify groups that match the spatial structure.
 If no spatial patterns is present in the data, our method did not identify any partitioning and standard clustering algorithms, which do not consider the spatial aspect, should be preferred in these circumstances.

 Whereas classical analysis consists in describing the main structures of ecological networks by summary statistics, our method is based on a model-based framework that assumes and estimates a distribution for interaction data. This approach is already used to analyze genomics data but it has been rarely applied to ecological data \citep[but see][]{pic09}. We demonstrated that this framework is very promising as it allows to include easily spatial constraints using a global model for the full network. It can be extended to build more complex models by considering covariates measured either on the nodes or the edges of the network (e.g., environmental variables). Lastly, different distributions can be modeled so that the method is applicable to a wide-variety of interaction data. For instance, we could assume a multivariate distribution to handle multi-layer ecological networks that record different types of interaction between entities \citep{kef12}.

%%%%%%%%%%%%%%%%%%%%%%%%%%%%%%%%%%%%%%%%%%%%%%%%%%%%%%%%%%%%%%%%%%%%%%%%%%%%%%
%%%%%%%%%%%%%%%%%%%%%%%%%%%%%%%%%%%%%%%%%%%%%%%%%%%%%%%%%%%%%%%%%%%%%%%%%%%%%%
%%%%%%%%%%%%%%%%%%%%%%%%%%%%%%%%%%%%%%%%%%%%%%%%%%%%%%%%%%%%%%%%%%%%%%%%%%%%%%
\section*{Acknowledgements}
The authors would like to thank Sophie Arnaud-Haond and Pierre Legendre for providing access to the hydrothermal fields dataset.

%%%%%%%%%%%%%%%%%%%%%%%%%%%%%%%%%%%%%%%%%%%%%%%%%%%%%%%%%%%%%%%%%%%%%%%%%%%%%%
%%%%%%%%%%%%%%%%%%%%%%%%%%%%%%%%%%%%%%%%%%%%%%%%%%%%%%%%%%%%%%%%%%%%%%%%%%%%%%
\section*{Appendix}
The fixed point algorithm arises when considering:
\begin{eqnarray*}
\frac{\partial}{\partial \tau_{iq}} \left(\mathcal{Q}(\bgamma, \bgamma^{[h]}) - \lambda \times \mathbb{E}_{\bgamma^{[h]}} \left\{\text{pen}(\bZ;\bLx)|\bY\right\} \right) & = & 
-\log\ \tau_{iq} -1 + \log\alpha_q +  \sum_{ji,\ell} \tau_{j \ell}\ \log f(Y_{ij}; \theta_{q\ell})  \\
&-& 2 \lambda\ \sum_{ij} X_{ij}(\tau_{iq}- \tau_{jq})  + L_i,
\end{eqnarray*}
where $L_i$ is the Lagrange multiplier that ensures the constraint $\sum_q \tau_{iq}=1$. Then the optimal parameter $\widehat{\tau}_{iq}$ satisfies:
$$
\widehat{\tau}_{iq}  \propto \alpha_q \left( \prod_{j \neq i} \prod_{\ell} f(Y_{ij}; \theta_{q \ell})^{\widehat{\tau}_{jl}} \right) \exp \left(-2 \lambda\ \sum_{i,j}  X_{ij}(\widehat{\tau}_{iq}- \widehat{\tau}_{jq}) \right)
$$

%%%%%%%%%%%%%%%%%%%%%%%%%%%%%%%%%%%%%%%%%%%%%%%%%%%%%%%%%%%%%%%%%%%%%%%%%%%%%%
%\section*{References}
\bibliographystyle{apalike}
\bibliography{mee_miele}

%%%%%%%%%%%%%%%%%%%%%%%%%%%%%%%%%%%%%%%%%%%%%%%%%%%%%%%%%%%%%%%%%%%%%%%%%%%%%%
%\end{multicols}

\newpage
\newpage
\newpage
\clearpage
\newpage

\end{document}